\documentstyle[preprint,aps]{revtex}
\draft
\begin{document}               
\title{
Condensed vortex ground states of rotating Bose-Einstein condensate
in harmonic atomic trap
}
\author{M. S. Hussein$^1$ and  O.K. Vorov$^{1,2}$
}
\address{
$^1$Instituto de Fisica, Universidade de Sao Paulo 
CP 66318,  05315-970,  
Sao Paulo, SP, Brasil 
\\
$^2$GANIL, BP 5027, F-14076,   Caen, Cedex 5, France
}
\date{24 November 2001}
\maketitle
\begin{abstract}
We study a system of $N$ Bose atoms trapped 
by a symmetric harmonic potential, 
interacting
via weak central forces. 
Considering the ground state of the rotating system
as a function of the two conserved quantities,
the total angular momentum and its collective component,
we develop 
an algebraic
approach
to derive exact wave functions and energies of these 
ground states.
We describe a broad class of the 
interactions for which these results 
are valid. This universality class is defined by simple
integral condition on the potential. 
Most of the potentials of practical interest   
which have pronounced repulsive component
belong to this universality class.
\end{abstract}

\vspace{30mm}

34 pages, 10 figures, no tables

\newpage

\section{Introduction}
\label{sec:level1}

Recent progress in creation of magnetic traps for cold atoms
made possible the study of the effects of quantum degeneracy 
in systems of finite numbers of interacting identical
particles  \cite{ANDERSON}.
One of the most interesting phenomena is the Bose-Einstein
condensation, which can be studied in different coupling regimes
in such dilute atomic gas systems. 
Theoretical studies of such systems of weakly interacting 
bosons, confined by 
parabolic potential, have become very active
\cite{nature,Wilkin,Mottelson,BP,REVIEW}.

Response of such system to rotation and the onset of vorticity
in the condensate
are among the most interesting questions 
attracting attention of both experimentalists  and theorists
\cite{nature,Wilkin,Mottelson,BP,REVIEW,Matthews,MarzlinZhangWright,A}.
In this connection, it is very important to study 
the structure 
and the spectrum of
the ground states of rotating condensates at a given angular momentum,
called the {\it yrast states}
\cite{nature,Wilkin,Mottelson,BP},
see 
Fig. 1.
The notion of yrast states is borrowed from nuclear physics,
where the states of many-body system having highest spins
at a given excitation energy play also a special role \cite{BM}.   

A particular problem of wide recent interest
\cite{Wilkin,Mottelson,BP}
arises
in the so-called weak coupling limit,
when the repulsive interactions between bosonic atoms 
can be considered
weak as compared to the spacing between different
oscillator levels in the harmonic trap.
This limit 
is expected 
\cite{nature,Wilkin,Mottelson}
to be reached in future experiments.
In particular, an interesting possibility to study 
different coupling regimes
is related to using the Feshbach resonance 
\cite{FESHBACH,STRENTHmanipulating}
to vary the strength of the effective interatomic interactions. 
From the theoretical viewpoint, 
the advantage of the weak coupling limit is that it 
allows analytical treatment of the problem.

We will consider here $N$ spinless Bose atoms 
in a spherically symmetric 
harmonic trap, assuming the interactions to be weak.
It is sufficient to consider the two-dimensional case,
because
the three-dimensional case can be reduced to 
the former
in the weak coupling limit, as was 
shown in Refs.\cite{Mottelson,BP}.
It is expedient to start with 
the case of noninteracting particles.
The noninteracting system 
has equidistant spectrum 
$\hbar\omega n$  
where $\omega$ is the trapping oscillator frequency 
and $n$ an integer.
Each level is degenerate, and the degeneracy
grows exponentially with $n$ for $n-N\gg1$ \cite{Mottelson}. 
This degeneracy is related to the number of 
ways to distribute the total 
energy $\hbar\omega n$
among the Bose
particles.
The short-range interactions $V(r)$ 
between the 
atoms are assumed weak 
in the sense that 
hoppings
between different
$\hbar\omega n$
levels can be neglected,
\begin{eqnarray} \label{WEAKcoupling}
N\langle V \rangle \ll \hbar\omega, 
\end{eqnarray}
where $\langle V \rangle$ is a typical matrix element 
of the interaction\cite{Mottelson}. 
The problem is therefore to find a nonperturbative
solution for the highly degenerate states at a single level 
$\hbar\omega n$, which is similar in spirit to the problem of
the lowest
Landau level for the electrons in high magnetic field,
which arises in the theory of 
the fractional quantum Hall effect \cite{Laughlin,Trugman-Kivelson}
or to the problem of 
compound
states in an
atomic nucleus.
The yrast states are those with minimal energy at given angular
momentum, $L$.
This is illustrated in 
Fig. 2, 
where the spectrum
patterns are shown for the cases of zero and finite interaction.
  
As is usually the case for interacting many-body systems, 
the evaluation of the exact 
ground state is a prohibitive task,
even with the simplification introduced by the weak coupling limit
\cite{Wilkin,Mottelson,BP}, 
\cite{HV1HV2,NOV1,NOV2,NOV3,dob1,dob2,dob3,HVprl,physicaB,HV3,praCOM}.
The yrast states in the case of {\it attractive} $\delta$-forces
have been found analytically by  Wilkin {\it et al.}\cite{Wilkin}.
Later, these results have been shown valid for a broad class of attractive
interactions in Ref.\cite{praCOM}.
The case of repulsive interaction is more difficult to analyze
\cite{Wilkin,Mottelson,BP,HV1HV2,NOV1,NOV2,NOV3,dob1,dob2,dob3}.
One of the first important results obtained 
for the
repulsive forces
was that of 
Bertsch and Papenbrock\cite{BP}:
these authors diagonalized the
repulsive $\delta$-interaction numerically
and suggested analytical formulae for the wave functions and
energies of the yrast states.
Later, it was shown analytically by several authors
using various methods
\cite{NOV3,NOV1,NOV2,HV1HV2,dob1},
that the states of the form of Bertsch and Papenbrock
are indeed eigenstates of the Hamiltonian. 
Only recently, it was  shown that these
states indeed correspond to minimum energy
\cite{HVprl},\cite{physicaB}.

In this work, we provide 
rigorous 
analytical solution for the yrast states.
In fact, we consider a more general problem \cite{HVprl},
of finding the ground state
as a function of two quantum numbers,
the total angular momentum, $L$,
and
the angular momentum of internal excitations (we discuss this
quantity in the following sections). 
Such  ``generalized yrast states'' include the
usual yrast states as a limiting case. 
These solutions are valid in the region $L \le N$.
Physical interpretation of the results is quite transparent.
In fact,
the form of the yrast wave functions which
was drawn from numerics \cite{BP} for the case of $\delta$-interaction,
turns out to be valid 
for a broad class of repulsive interactions.
The universality class of such interactions
[which includes, but is not limited to, delta-forces, 
Gaussian forces with arbitrary range,
Coulomb ($1/r$) and log-Coulomb [$log(r)$] forces]
is described by an explicit sufficiency condition.

The structure of the paper is the following.
In section II, we 
present detailed formulation of the problem in the weak coupling
limit,
introducing the relevant ``partition subspace'' 
and the basic ingredients of the 
following consideration.
In section III, we derive the operator expansion for the Hamiltonian
projected onto the partition space. 
In section IV, we discuss the additional conserved quantum number,
``seniority'',
which can be interpreted as a collective contribution to the 
angular momentum,
and introduce the notion of generalized yrast states.
The method of 
algebraic 
decomposition of the Hamiltonian 
is 
introduced and discussed in section V.
Its application to the present problem is developed 
in section VI. 
Section VII is devoted to study the properties of 
eigenvalues of the ``perturbation'' defined in the section  VI.
The results for the generalized yrast states are presented,
in general form, in section VIII.
In section IV, we consider various examples of the interaction
potentials and discuss the results.
Section X is devoted to detailed discussion of 
the applicability condition
and of the corresponding universality class of attractive potentials.
Section XI summarizes the work.

\section{The partition subspace}
\label{sec:level2}

In the weak coupling limit discussed in the previous section,
the problem splits onto series of independent problems for
each value of the total angular momentum $L$.
At given angular momentum $L$,
the Hamiltonian 
($\hbar=m=1$) 
is the sum
\begin{eqnarray}\label{ham}
\tilde{H}= 
\omega \tilde{H}_0
+ \tilde{V},
\end{eqnarray}
where the first term 
(the c-number equal to the energy of degenerate level)
comes from the noninteracting Hamiltonian 
in the parabolic trap,
\begin{eqnarray}\label{H0}
\omega H_0 = \sum_i^{N}
\left(\frac{\vec{p}_i^2}{2}+\frac{\omega^2}{2}\vec{r}_i^2 \right)
\qquad \rightarrow \qquad \omega \tilde{H}_0,
\end{eqnarray}
with $\vec{p}_i$ and $\vec{r}_i$ denoting the
momentum and coordinate of the $i$-th particle. 
Hereafter, 
we set $\omega\equiv 1$ for brevity.
The second, nontrivial, term $\tilde{V}$ stems
from the two-body interaction 
\begin{eqnarray}\label{INTERACTION}
V=\sum\limits_{i>j} V(r_{ij}) \qquad \rightarrow \qquad \tilde{V} ,
\end{eqnarray}
projected onto the single-level problem,
which is worked out below.

In two dimensional problems involving harmonic 
potentials,
it is convenient to use 
the complex variable $z=x + iy$ and the conjugated $z^*=x - iy$
instead of the position vector $\vec{r}=(x,y)$.
Introducing the notations 
$\frac{\partial}{\partial z} = \frac{1}{2}(\frac{ \partial}{ \partial x}
- i\frac{ \partial}{ \partial y})$
and  
$\frac{\partial}{\partial z^*} = 
\frac{1}{2}(\frac{ \partial}{ \partial x}
+ i\frac{ \partial}{ \partial y})$,
it is convenient to employ
the tetrad of 
ladder
operators
$a^{+}$, $a$, $b^{+}$ and  $b$
\begin{eqnarray} \label{a-akrest}
a^{+} =  \frac{z}{2}- \frac{\partial}{\partial z^*}, \quad 
b^{+} =  \frac{z^*}{2}- \frac{\partial}{\partial z} , \quad 
\nonumber\\
a = (a^{+})^{\dagger} =  
\frac{z^*}{2}+ \frac{\partial}{\partial z}   , \quad
b = (b^{+})^{\dagger}  =  \frac{z}{2}+ \frac{\partial}{\partial z^*}    , 
\end{eqnarray}
for each
particle. For brevity, the particle markers are 
suppressed in Eqs.(\ref{a-akrest}), dagger denotes Hermitean
conjugation. 
The role of these 
operator is to raise (lower) 
the powers of $z_i$ and $z^*_j$ in the 
preexponentials of many-body wave functions, 
which are all polynomials times the Gaussian factor  
$|0\rangle=exp(-1/2\sum|z_k|^2)$.
Thus, we have
\begin{eqnarray}\label{aGROUND}
z_i|0\rangle = a^+_i|0\rangle, \quad
z^*_j|0\rangle = b^+_j|0\rangle, 
 \nonumber\\
\quad a_i|0\rangle=0, \quad 
b_i|0\rangle=0, \qquad etc .
\end{eqnarray}

The only nonzero commutators between (\ref{a-akrest}) are 
given by
\begin{eqnarray}\label{COMab}
[a_i,a^{+}_j] = [b_i,b^{+}_j] = \delta_{ij} .
\end{eqnarray}

In two dimensions, 
the total angular momentum 
is 
the difference 
\begin{eqnarray}
L=L_+ -L_-
\end{eqnarray}
of numbers 
of ``up'' and ``down'' quanta,
$L_+=\sum_1^N a^{+}_ka_k$ 
and 
$L_-=\sum_1^N b^{+}_kb_k$, respectively.
At the same time, the noninteracting Hamiltonian on the left 
hand side of (\ref{H0}) is  
given by 
\begin{eqnarray}
H_0=L_+ +L_-+N .
\end{eqnarray}
By definition, the yrast states have minimum energy at given $L$.
They must therefore belong to the subspace
with 
\begin{eqnarray}\label{YRAST-CONDITION}
L_{-} = 0 , \qquad L_{+} = L .
\end{eqnarray}
The first term in (\ref{H0}) is therefore 
reduced to a constant given by
\begin{eqnarray}
\tilde{H}_0 = N + L .
\end{eqnarray}
The  subspace defined by (\ref{YRAST-CONDITION}) is
spanned 
by the homogeneous symmetric polynomials of degree $L$,
$poly^{L}_S(z_i)$ 
which do not involve conjugated $z^*$'s at all,
\begin{eqnarray}\label{polyPARTITION}
\Psi(L) = poly^{L}_S(a^{+}_i) |0\rangle
\end{eqnarray}
The orthogonal basis of these symmetric 
polynomials can be constructed as follows 
\cite{symmetricFUNCTIONS}.
Consider the monomials
\begin{equation} \label{MONO}
m \equiv 
\{ {l_1},{l_2},...,{l_N} \}
\equiv 
z_1^{l_1}z_2^{l_2}...z_N^{l_N} 
\end{equation}
corresponding to partition of integer $L$,
$\sum\limits_n l_n = L$.
The states $m|0\rangle$ with different sets
$\{ {l_1},{l_2},...,{l_N} \}$
are mutually orthogonal.
The basis symmetric polynomials \cite{symmetricFUNCTIONS}
are given by symmetrized 
linear combinations of (\ref{MONO})
\begin{equation} \label{PARTITION}
[{l_1},{l_2},...,{l_N}] \equiv 
P_S
m, 
\end{equation}
denoted by corresponding ordered partition 
$[{l_1},{l_2},...,{l_N}]$ with 
${l_1} \ge {l_2} \ge ... \ge {l_N}$. 
Here, 
the operator of symmetrization 
$P_S$ 
is the standard
sum over all the permutations.
In the case $L \le N$ which we are interested in here,
the number of the basis states 
(\ref{PARTITION})
is given by 
``the number of
unrestricted
partitions'' $[{l_1},{l_2},...,{l_N}]$
of positive integer $L$ \cite{ABR}. 
At $L\leq N$, 
this number
does not depend on $N$; hereafter, it is denoted by $p(L)$.
For example, at $N = L = 3$ 
the complete basis (\ref{PARTITION}) 
of the partition space (\ref{polyPARTITION})
is spanned by
the following states 
\begin{eqnarray} \label{example}
{[} 1,1,1 {]} = z_1 z_2 z_3 = |0,3,0,0...\rangle , 
\qquad \qquad \qquad \qquad \qquad  \qquad \qquad 
\nonumber\\ 
{[} 2, 1, 0 {]} = z_1^2 z_2 
+ z_1 z_2^2 +z_2^2 z_3 +z_2 z_3^2 + z_3^2 z_1 +z_3 z_1^2 
= |1,1,1,0...\rangle , 
\nonumber\\
{[} 3, 0, 0 {]}=z_1^3 +z_2^3 +z_3^3
= |2,0,0,1...\rangle ,
\qquad \qquad \qquad \qquad \qquad  \qquad 
\end{eqnarray}
which correspond
to the number of partitions $p(3)=3$ 
(normalization factors and the Gaussian, $|0\rangle$,
are suppressed for brevity). 
In Eqs.(\ref{example}), the 
right hand side
of each line gives
interpretation of many-body state in terms of 
the single-particle occupation 
numbers of the oscillator states $|n_0, n_1, n_2,...\rangle$
with $n_i$ the number of Bose particles in the state with $i$
oscillator
quanta.

The numbers of partitions, $p(L)$, can be computed from 
the generating function, using the formula
\begin{displaymath}
\sum\limits_{L=0}^{\infty} p(L) t^L = 
\prod\limits_{k=1}^{\infty}\frac{1}{1-t^k} .
\end{displaymath}
An explicit (however involved) expression for $p(L)$ 
is given in
\cite{ABR}. 
For small $L$, the values of $p(L)$  are
\begin{eqnarray}\label{p-L}
p(2)=2, \qquad
p(3)=3, \qquad
p(4)=5, \qquad
p(5)=7, \qquad
\nonumber\\
p(6)=11, \qquad
p(7)=15, \qquad
p(8)=22, \qquad ...
\end{eqnarray}
At $L$ large, the number of 
states grows exponentially \cite{ABR} 
\begin{equation} \label{asympt}
p(L \gg 1) \propto  e^{L^{\frac{1}{2}}},
\end{equation}
as is very typical for many-body systems \cite{BM}.

The problem of finding the spectrum consists of
diagonalization of the interaction, the second term in (\ref{ham}),
in the ``partition subspace'' (\ref{polyPARTITION},\ref{PARTITION});
the eigenvalues of (\ref{ham}) have the form 
\begin{equation}
{\cal E} = N + L + E,
\end{equation}
where $E$ is the interaction contribution.
It should be noted that this term can not be found with any kind
of perturbation theory, as it comes from diagonalization
of the interaction within the subspace of degenerate states.

In this paper, we restrict ourselves to the case $L \leq N$ only. 
At $L > N$, the whole situation is changed drastically. 
First, the 
number of basis states is not given by $p(L)$ anymore
and it depends on both $L$ and $N$.
For example, the simplest states with partition $[1,1,1...1]$, which
play 
the role of generating functions for the 
ground states at $L \leq N$ (see section V),
can not be constructed at $L > N$.
It is therefore reasonable to expect something similar to
``phase transition'' at $L=N$. In the earlier studies 
\cite{BP}, the signatures of this phase transition
have been observed in numerical simulations.

\section{The interaction within the partition subspace.
``Operator expansion''}
\label{sec:level4}

In this section, we work out a convenient representation for
the $\tilde{V}$, Cf. (\ref{INTERACTION}),
the interaction $V$ projected onto the ``partition space'' which
was described in the previous section.
Using condition Eq.(\ref{YRAST-CONDITION}),
the projected interaction $\tilde{V}$
in Eq.(\ref{ham})
can be written \cite{HV1HV2,HVprl} as
\begin{eqnarray}\label{projection}
\tilde{V} = \sum\limits_{L=-\infty}^{\infty}
P^-_{0} P^+_{L} V P^+_{L} P^-_{0} ,
\end{eqnarray}
where we use the standard number-of-quanta projector
\begin{eqnarray}\label{PROJECTOR}
P^+_{L} = \frac{1}{2\pi}
\int_0^{2\pi} d \alpha e^{i \alpha \left(L-\sum_ia^{+}_ia_i\right)},
\end{eqnarray}
and the similar expression for $P^-_{L}$ with substitutions
$a^{+}_i \rightarrow b^{+}_i$ and
$a_i \rightarrow b_i$.
Due to construction, the transitions between different $L$-sectors
are suppressed in (\ref{projection}), and the resulting effective
interaction
(\ref{projection}) 
applies equally to every $L$-sector.

With the help of (\ref{PROJECTOR}), Eq.(\ref{projection})
can be cast in the form
\begin{eqnarray}\label{projection1}
\tilde{V} = \frac{1}{2\pi}\int\limits_{-\infty}^{\infty}
d\alpha
P^-_{0} e^{i\alpha\sum_ia^{+}_ia_i}  V 
e^{-i\alpha\sum_ia^{+}_ia_i} P^-_{0} .
\end{eqnarray}
In order to evaluate (\ref{projection1}), 
we use the Fourier representation of the interaction in the right
hand side of (\ref{projection1}),
\begin{eqnarray}\label{fourier} 
V = 
\sum\limits_{i>j}
\int\limits_{-\infty}^{\infty}
d q_x
\int\limits_{-\infty}^{\infty}
d q_y
e^{ i\left[ q_- 
(a^{+}_{ij}+b_{ij})+ q_+ (b^{+}_{ij}+a_{ij})
\right]} 
V_q 
\end{eqnarray}
where 
\begin{displaymath}
V_q= \frac{1}{2\pi}
\int_{0}^{\infty} rdr
J_0\left(r\sqrt{q_x^2+q_y^2}\right)
V(r)
\end{displaymath}
is the two-dimensional Fourier transform of the central 
potential $V(r)$. 
Here, $J_0$ is the Bessel function \cite{ABR} and
notation
$q$$_{\pm}$$=$$($$q$$_x$$\pm$$i$$q$$_y$$)$$/$$\sqrt{2}$
is introduced.
The two-particle operator combinations 
\begin{eqnarray}\label{aa} 
a^{+}_{ij}\equiv\frac{1}{\sqrt{2}}(a^{+}_{i}-a^{+}_{j}), \quad 
b^{+}_{ij}\equiv\frac{1}{\sqrt{2}}(b^{+}_{i}-b^{+}_{j}), \quad 
\nonumber\\
a^{+}_{ij}\equiv (a^{+}_{ij})^{\dagger}, \quad 
b^{+}_{ij}\equiv (b^{+}_{ij})^{\dagger}
\end{eqnarray}
came from resolving 
$x_{i},y_{i}$ and $x_{j},y_{j}$ 
from (\ref{a-akrest}).
Substituting (\ref{fourier}) in (\ref{projection}), we make use of 
Baker-Hausdorf relation 
\begin{displaymath}
e^{a^{+} - a}
= e^{\frac{[a^{+},a]}{2}} e^{a^{+}}e^{-a},
\end{displaymath}
which is valid for any pair of boson operators $a^{+}$ and $a$. 
Expanding (\ref{fourier}) to powers of $q^2$ and evaluating 
integrals term by term, we use the relation
\begin{displaymath}
\int_0^{\infty} dq e^{-q^2/2}q^{2n+1}J_0(qr)=2^nn!M(n+1,1,-r^2/2)
\end{displaymath}  
where  $M(\mu,\nu,x)$ is the Kummer confluent hypergeometric
function \cite{ABR}.
Proceeding in this manner, 
we obtain 
expansion of $\tilde{V}$
\begin{eqnarray}  \label{OPE0}
\tilde{V} = \sum\limits_{k=0}^{2[L/2]}(-1)^ks_kB^{k} =
\frac{N(N-1)}{2} - s_1 B^{1} +... + 
\end{eqnarray}
in terms of the two-particle normal-ordered operators 
defined as
\begin{equation}
B^k = \sum\limits_{j>i}^{N} B^k_{ij} ,
\qquad
B^k_{ij} = a^{\dagger k}_{ij} a_{ij}^k .
\end{equation}
In Eq.(\ref{OPE0}), $[q]$ denotes integer part of real 
number $q$.
The ``strength parameters'', $s_k$, are 
related to the potential, $V(r)$,
via the integrals
\begin{eqnarray}  \label{OPE2}
s_k =  
\frac{1}{k!} \int_0^{\infty} dt M(k+1,1,-t) V(\sqrt{2t}).
\end{eqnarray}
with $M$ the Kummer function \cite{ABR}.
For $k$ integer,
the Kummer function 
can be expressed in terms of the Laguerre polynomials $L^0_n$ 
\begin{displaymath}
M(k+1,1,-t) = e^{-t} L^0_n (t) .
\end{displaymath}
In the expansion (\ref{OPE0}),
the highest possible order $k$ of the operators
$B^k$ is $L$ for $L$ even,
and $L-1$ for $L$ odd.
In the lowest order term, no operators are involved, and 
we have $B^0 \equiv N (N - 1 ) / 2$.

Eq.(\ref{OPE0}) can be written in another convenient form
\begin{eqnarray}  \label{OPE}
\tilde{V} = 
\sum\limits_{k=0}^{[L/2]} V_k,
\nonumber\\
\qquad
V_k\equiv s_{2k}(B^{2k}-B^{2k-1})+
(s_{2k+2}-s_{2k+1})B^{2k+1}.
\end{eqnarray}
Here, $B^{-1} \equiv 0$.
This second form of the expansion 
contains operator structures collected in the way
convenient to apply the method of 
``algebraic decomposition''
described in the following sections.

The operator expansion
(\ref{OPE0},\ref{OPE}) is universal,
while the particular shape of the
potential $V(r)$ is described by 
the integrals with Kummer function $M$\cite{ABR}.
We should stress that
expansion (\ref{OPE}) is exact for any interaction
$V(r)$ whose moments $s_k$ are finite\cite{HV1HV2,HV3}.

\section{The quantum number ``seniority'',
generalized yrast states and correspondence rules}
\label{sec:level3}

In this section, we discuss an extra conserved quantum number
the system enjoys, in addition to the energy and the total
angular momentum.
\cite{Wilkin,Mottelson}. This quantity, 
which is very important
for classification of states,
is reminiscent to the ``center-of mass'' mode in 
nuclear problems \cite{BM}.
Here, it can be interpreted as
collective contribution to the total angular momentum $L$.
Indeed, the  pair of the mutually conjugated collective ladder operators 
\begin{eqnarray}
{\cal A}^{+}=\sum_{i=1}^{N}\frac{a^{+}_i}{\sqrt{N}}, \qquad
{\cal A}=\sum_{i=1}^{N}\frac{a_i}{\sqrt{N}}, \qquad
[{\cal A}, {\cal A}^{+}]=1 ,
\end{eqnarray}
commute 
with any two-body combinations in (\ref{OPE}) as well as with the
angular momentum,
\begin{eqnarray}\label{COMMUTATION}
[ {\cal A}, a^+_{ij} ] = 0, \qquad [ {\cal A}, L_+ ] = 0, 
\qquad [ {\cal A}, L_- ] = 0,
\end{eqnarray}
and they therefore commute with both terms in the 
Hamiltonian
$\tilde{H}$ (\ref{ham}),
thus we have $[{\cal A},\tilde{H}]=0$.
The number 
of collective quanta, ${\cal A}^{+}{\cal A}$,
is therefore a conserved quantity.  
The mutual eigenfunctions of the triad of operators 
$\tilde{H}$, $L$ and 
${\cal A}^{+}{\cal A}=v$ 
can be found in following factorized form 
\begin{eqnarray} \label{seniorityBASIS}
\Psi_k(L,v) = \qquad Z^{v} \quad poly^{L-v}_S
(\tilde{z}_i) \quad
e^{-\sum\frac{|z_k|^2}{2}},
\end{eqnarray}
where $\tilde{z}_i = z_i- Z$ 
and 
\begin{eqnarray}
Z \equiv \sum\limits_{i=1}^N z_i / N
\end{eqnarray}
is the collective variable.
The additional index $k$ stands to distinguish between the
different states in the same $(L,v)$-sector.

In the state (\ref{seniorityBASIS}), 
the degree of the pre-exponential polynomial, $L$,
which is 
the
total angular momentum, 
is redistributed 
between internal excitations, $J$$=$$L$$-$$v$, and
the contribution due to the collective motion, $v$,
which we call {\it seniority} for brevity.

The energies of the states (\ref{seniorityBASIS}),
i.e., the eigenvalues of (\ref{ham}) have the form
\begin{eqnarray}
{\cal E}_k(L,v) = N + L + E_k(L,v)
\end{eqnarray}
where $E_k(L,v)$, the interaction energy, comes from
diagonalization of $\tilde{V}$.
It is therefore meaningful to consider the ground 
state as a function of 
both
$L$ and $v$, as illustrated
Fig. 3.
We call these states, 
\begin{eqnarray}\label{generalizedYRAST}
\Psi_0(L,v), \qquad {\cal E}_0(L,v) = min_{k} \{ {\cal E}_k(L,v) \} 
\end{eqnarray}
with minimum energy  at fixed
values of $L$ and $v$ the ``generalized yrast states''
\cite{HVprl},\cite{physicaB}.
The usual yrast states are those among the (\ref{generalizedYRAST}),
which minimize the energy ${\cal E}_0(L,v)$ with respect to 
the seniority,  
\begin{equation} \label{YRASTmin}
{\cal E}_{yrast}(L) = min_v \{ {\cal E}_0(L,v) \} .
\end{equation}
At given $L$, one has exactly $L$ allowed values 
of the seniority, they are
\begin{equation} \label{allowed}
v = 0 , 1 , 2 , ... , L-2 , L.
\end{equation}
The value
$v=L-1$ is excluded  
because the symmetric degree $J=1$ polynomial in variables 
$z_i-Z$ is evidently reduced to zero,
\begin{displaymath}
poly^{1}_S(z_i-Z)=\sum\limits_{i=1}^{N}(z_i-Z)\equiv 0.
\end{displaymath}

The states with definite seniority 
of type (\ref{seniorityBASIS}), can be obtained 
by applying the seniority projector 
[Cf. Eq.(\ref{PROJECTOR})]
\begin{eqnarray}\label{SENIORITY-PROJECTOR}
{\cal P}_{v} = \frac{1}{2\pi}
\int_0^{2\pi} d \phi e^{i \phi (v-{\cal A}^{+}{\cal A})}
\end{eqnarray}
to the states of type (\ref{PARTITION}).

The quantum number seniority helps to establish very useful
``corresponding rules'' between the states in different
$(L,v)$-sectors and to relate their spectra.
Indeed, by virtue of (\ref{projection}),(\ref{SENIORITY-PROJECTOR}) and 
(\ref{COMMUTATION}),
we can write
\begin{eqnarray}\label{correspondence1} 
\tilde{V} \Psi_k(L+1,v+1) = \tilde{V} {\cal A}^{+}\Psi_k(L,v) =
\nonumber\\
= {\cal A}^{+}\tilde{V}  \Psi_k(L,v) = E_k(L,v) {\cal A}^{+}
\Psi_k(L,v) . 
\end{eqnarray}
Eq.(\ref{correspondence1}) means the following: let 
$ \{ \Psi_k (L,v) \}$ be 
the $p(L)$ normalized eigenstates for the Hamiltonian (\ref{ham})
in the sector $L$. Then exactly $p(L)$ eigenstates in the
sector $L+1$ (having nonzero $v$!) can be obtained simply
by applying the collective ladder operator to the states
in $L$ sector,
\begin{eqnarray}\label{correspondence2} 
\Psi_k(L+1,v+1) = \left(\frac{v}{v+1}\right)^{1/2} {\cal A}^{+}\Psi_k(L,v), 
\end{eqnarray}
while their interaction energies will be the same,
\begin{eqnarray}\label{correspondence3} 
E_k(L+1,v+1) = E_k(L,v) . 
\end{eqnarray}
Similar relationships appear in the fermionic problem of the
fractional quantum Hall effect \cite{Trugman-Kivelson}.

As the total angular momentum grows from $L-1$ to $L$,
the only new structures in the wave functions appear 
in the {\bf seniority zero sector},
where we have new states of total number equal to
\begin{eqnarray}\label{correspondence4}
g_0(L) = p(L) - p(L-1) ,
\end{eqnarray}
They must be obtained by diagonalization of $\tilde{V}$
in the sector $v=0$.

Using (\ref{correspondence4}) recursively from $v$ to $v+1$,
we calculate the total numbers of states, $g_v(L)$,  
in the $(L,v)$-sectors 
\begin{eqnarray}\label{correspondence5}
g_v(L) = p(L-v) - p(L-v-1) \qquad 
for \qquad v\leq L - 2 
\nonumber\\
and \qquad g_{v=L}(L)=1.
\end{eqnarray}
These important relations will be very usefull in the analysis
given in the following sections.

\section{Algebraic decomposition of the Hamiltonian}
\label{sec:level5}

Regular methods to obtain
exact ground state 
without solving the whole spectrum are not available.
We use the approach 
\cite{HV3} which we loosely nicknamed 
``algebraic decomposition''
\cite{SUSYprimer}.
Suppose that the Hamiltonian 
can be written as a sum, 
\begin{eqnarray}\label{SUSY1}
\tilde{V}= V_0 + V_S,
\end{eqnarray}
such that 
\\
{\bf (a)} the first term, $V_0$, 
is simple, and one can
find out its ground state $|0)$
with eigenvalue ${\cal E}_{min}$, possibly degenerate. 
If the second term, $V_S$,  
has
the two 
properties:

{\bf (b)}
$V_S$ annihilates the state $|0)$, 
so $V_S|0)=0$,

{\bf (c)} 
$V_S$ is {\it non-negative definite}, $V_S\geq 0$, (it does not have
negative eigenvalues),
\\
then 
the state $|0)$ will still be the ground 
state 
for the full Hamiltonian $V_0+V_S$, with 
the same eigenvalue ${\cal E}_{min}$.
Indeed, 
{\bf (b)} implies that $|0)$ is
an eigenstate for the sum $V_0+V_S$ with its eigenvalue
intact. 
As one knows from linear algebra, 
if an Hermitean 
operator 
$V_0$ is perturbed
by {\it a non-negative definite} Hermitean operator $V_S$, the 
eigenvalues can only increase (see, e.g. \cite{BOOK-MATRICES}).
This means that the states other than $|0)$
can gain energy [Cf. {\bf (b)}].
As $|0)$ is already the ground state for $V_0$, it
will be the same for $V_0+V_S$
\cite{COMMENTdegeneracy}.

A trivial but helpful example is the one-dimensional
linear oscillator 
with the Hamiltonian $V=1/2+a^+a$. The state
$|0)$, which obeys $a|0)=0$, is ``the ground state'' for
$V_0\equiv 1/2$ because any other state $|n)$ has the same
eigenvalue $1/2$. The non-negative definite 
operator $V_S=a^+a$, $V_S\geq0$ can be considered as 
``perturbation''.
It removes 
this degeneracy and leaves $|0)$ with lowest eigenvalue of
the total Hamiltonian
$V$ \cite{COMMENTdegeneracy}.

The same arguments apply to the case with
additional conserved quantum number $v$, such as
$[v , V_0]=[v,V_S]=0$.
In the above scheme, the single state $|0)$ 
is replaced by the set of
states $|L,v)$,
each having the property {\bf (a)} and {\bf (b)} in their $v$-sectors.
This is illustrated in 
Fig. 4, 
where the spectrum 
of $V_0$ is taken degenerate in each $v$-sector.

\section{Algebraic triad}
\label{sec:level6}

The starting strategy is
to look 
for states annihilated by a part of $V$ 
which can be then shown to be non-negative definite operator.
In our case, the 
``algebraic triad''
$V_0$, $V_S$ and $|0)$
can be established by inspecting action of terms $V_k$
in (\ref{OPE}) on selected states of partition basis (\ref{PARTITION}).
We observe first, that
using the algebraic identity
\begin{displaymath}
N\sum_i^N a^+_i a_i = \frac{1}{2} \sum_{i,j}^N(a^+_i -a^+_j)(a_i-a_j)+
\sum_i^Na^+_i \sum_j ^Na_j ,
\end{displaymath}
one can cast the first term, $V_0$, of the operator expansion
(\ref{OPE})
in the following form,
\begin{eqnarray}\label{V0}
V_0=\frac{N}{2}[(N-1)s_0 -(L-{\cal A}^{+}{\cal A})(s_1 -s_2)].
\end{eqnarray}
We recall now that the operator ${\cal A}^{+}{\cal A}$ is diagonal 
in the seniority basis (\ref{seniorityBASIS}).
In this basis, the operator $V_0$ is therefore reduced to a
combination of quantum numbers, (\ref{V0}), 
with ${\cal A}^{+}{\cal A}=v$.
Next, one can see that  
the simplest basis state, $|L\rangle$, [Cf. Eq. (\ref{PARTITION})]
with partition $[1,1,...1]$
\footnote{An example of such states is given by the first 
line of (\ref{example}).}
\begin{eqnarray}\label{simplest}
|L\rangle = P_S a^+_1 a^+_2 ... a^+_L |0\rangle,
\end{eqnarray}
is annihilated by the remainder of the Hamiltonian (\ref{OPE}),
\begin{eqnarray}\label{simplest1}
(\tilde{V} - V_0) |L\rangle=0 .
\end{eqnarray}
In order to prove (\ref{simplest1}),
we use Eqs.(\ref{COMab},\ref{aa}) to evaluate the commutator
\begin{eqnarray}\label{co}
[a_{12}^2, a^+_1 a^+_2]=-2(a^{+}_{12}a_{12}+1).
\end{eqnarray}
From this equation, it follows that 
\begin{displaymath}
a_{12}^{k} a^+_1 a^+_2a^+_3...a^+_L|0\rangle=0
\end{displaymath}
for any $k > 2$, by virtue  of (\ref{aGROUND}).
Therefore, we have  
\begin{eqnarray}\label{zero1}
B^{k} | L \rangle = 0
\end{eqnarray} 
for any $k > 2$.
From the same commutator (\ref{co}), we have
\begin{displaymath}
[a^{+2}_{12}a_{12}^2, a^+_1 a^+_2]|0\rangle=
[ a^{+}_{12}a_{12}, a^+_1
a^+_2]|0\rangle=-a^{+2}_{12}|0\rangle ,
\end{displaymath}
and thus $(B^{2}-B^{1})|L\rangle=0$. This relation together 
with (\ref{zero1}) results in (\ref{simplest1}).

Equations (\ref{V0}) and (\ref{simplest1}) hint that  
the state $|L\rangle$ and the operator $V_0$ may constitute important
ingredients of the 
algebraic  decomposition,
Cf. Eq.(\ref{SUSY1})
described in the previous section.
However,  the state $|L\rangle$ does not have 
definite seniority and it can not be eigenstate of the Hamiltonian.
Nevertheless, as we will see, the $|L\rangle$ turns to be 
a generating function for states with definite seniority.
Indeed, substitution 
$z_i=\tilde{z}_i+Z$ [see (\ref{seniorityBASIS})]
transforms $|L\rangle$ to a sum
\begin{eqnarray}\label{multivortex2}
|L\rangle = 
\biggl[
P_S \tilde{z}_1 \tilde{z}_2 ...\tilde{z}_L
+...
+Z^{L-2}P_S\tilde{z}_1 \tilde{z}_2
+ Z^{L} 
\biggr]|0\rangle =
 \nonumber\\
 =
\sum\limits_{v=0}^{v=L(v\neq L-1)}  {\cal P}_v|L\rangle
\end{eqnarray}
of exactly $L$ 
states
of the form 
(\ref{seniorityBASIS}), 
each being the eigenvector 
of ${\cal A}^+{\cal A}$ and therefore of $V_0$, 
with the eigenvalue (\ref{V0}).
We notice that each $v$-sector is represented
by a single term in (\ref{multivortex2}),
identified with ${\cal P}_v|L\rangle$.

The spectrum of $V_0$ is very simple.
A schematic example is shown in 
Fig.4, 
left hand side.
It consists of $L$ 
equidistant (except $v$$\neq$$L$$-$$1$), 
$g_v(L)$-fold 
degenerate levels with 
energies given by
(\ref{V0}). The degeneracies are given by  
Eq.(\ref{correspondence5}).
Each 
$v$-level contains one and only one state
$|L,v)={\cal P}_v|L\rangle$ from the sum (\ref{multivortex2}).
Therefore, the set of 
states 
$|L,v)$
obey the criterion
{\bf (a)} with the operator $V_0$.
The property {\bf (b)} with $V$$_S$$\equiv$$\tilde{V}$$-$$V$$_0$
holds by virtue of (\ref{simplest1}).
In particular, {\bf (a)} together with {\bf (b)} mean that 
$|L,v)$
are the eigenvectors of $\tilde{V}$$=$$V$$_0$$+$$V$$_S$ 
with eigenvalues (\ref{V0}).
This holds for any interaction $V(r)$.

The ``algebraic decomposition'' with the triad 
$V_0$, $V_S$
and $|L,v)={\cal P}_v|L\rangle$
would be complete if we succeeded to prove
non-negative definiteness $V_S$$\geq$$0$ of 
the remainder of the Hamiltonian $\tilde{V}-V_0$,
criterion {\bf (c)}.
So far, we did not specify form of the interaction $V(r)$ in 
(\ref{OPE}).
We will study now general case and specify the 
class of potentials which have $V_S\geq 0$.

\section{Non-negative definiteness of 
``perturbation''}
\label{sec:level7}

We have to check signs of all the eigenvalues of $V_S$
in the partition space (\ref{PARTITION}). 
To avoid solving the whole spectrum 
in the space of symmetrized states,
we use the following
trick. 
By definition, the nonzero eigenvalues of 
$V_S$
in the space (\ref{PARTITION}) coincide
with the nonzero eigenvalues of $P_S V_S P_S$ in the 
{\bf full space of monomials} $m$ in (\ref{PARTITION}).
This latter space has dimensionality 
much higher than $p(L)$
and it includes wave functions of all possible symmetries,
including boson sector (fully symmetric), 
fermion sector (fully antisymmetric) {\it etc}.
In this extended space, the 
analysis of signs of eigenvalues 
is however crucially simplified, 
while the contributions from the symmetric sector can
be accurately separated.
Using the symmetry of $V_S$ under permutations of particles, 
we can write
\begin{eqnarray}\label{INERTIA}
P_S V_S P_S =  
P_S \sum\limits_{i>j}V_{S,ij}  P_S 
= 
\frac{N(N-1)}{2}
P_S 
V_{S,12}P_S 
\end{eqnarray}
where $V_{S,ij}$ is the contribution from pair of particles $i$$,j$
to $V_S$ [Cf.(\ref{OPE},\ref{OPE2},\ref{SUSY1})].
In order to see that 
\begin{equation}\label{susy3}
P_S V_S P_S \geq 0 
\end{equation} 
for a given interaction potential $V(r)$,
it is {\it sufficient}
to show that 
$V_{S,12}$$\geq$$0$,
because application of any 
projector 
$P_S$ 
from both sides in (\ref{INERTIA})
can add new zero eigenvalues, but can not add
negative eigenvalues.
This follows from the known ``inertia theorem'' of linear algebra
\cite{BOOK-MATRICES}.
Now, we study the eigenvalues of $V_{S,12}$.
Let $\pi_{12}$ be the operator of permutation of 
variables $1$ and $2$.
The triad 
\begin{displaymath}
T=\{\pi_{12}, \quad a^{+}_{12}   a_{12}, \quad V_{S,12} \}
\end{displaymath}
forms a set of mutually commuting operators. 
Indeed, $V_{S,12}$ 
is expressed 
in terms of $B^k_{12}=a^{+k}_{12}$$a^k_{12}$ 
[see (\ref{OPE})]. 
For fixed pair of particle indices, the operators
$a_{12}$ and  $a^+_{12}$ commute like Bose operators, 
$[ a_{12}, a^+_{12}] = 1$.
Consequently,
any operator $B^k_{12}$
can be expressed in terms of $a^+_{12}a_{12}$,
using the standard boson calculus formula
\begin{eqnarray} \label{calculus}
a^{+ k}_{12}a^k_{12}=
a^{+}_{12}a_{12}(a^{+}_{12}a_{12}-1)...
(a^{+}_{12}a_{12}-k+1). 
\end{eqnarray}
The triad $T$ is diagonalized simultaneously in the 
basis of monomials
$m\equiv  z_1^{l_1} z_2^{l_2} z_3^{l_3}... z_N^{l_N} $,
(\ref{PARTITION})
with the only substitutions 
$z_1$$\rightarrow$
$\frac{1}{\sqrt{2}}$$($$z_1$$-$$z_2$$)$,
$z_2$$\rightarrow$$\frac{1}{\sqrt{2}}$$($$z$$_1$$+$$z$$_2$$)$,
\begin{eqnarray}\label{NEWbasis}
\left[\frac{1}{\sqrt{2}}(z_1-z_2)\right]^{l_1} 
\left[\frac{1}{\sqrt{2}}(z_1+z_2)\right]^{l_2}
z_3^{l_3}... z_N^{l_N} |0\rangle . 
\end{eqnarray}
In this basis, the eigenvalues of the triad $T$
depend only on $l_1$ through the subfactor
$\biggl[$$\frac{1}{\sqrt{2}}$$($$z$$_1$$-$$z$$_2$$)$$\biggr]$$^{l_1}$
of the eigenvector 
(\ref{NEWbasis}).
These eigenvalues are given by 
\begin{eqnarray}\label{TRIAD}
T = \left\{ (-1)^{l_1} , \quad l_1, \quad \lambda_{l_1} \right\},
\end{eqnarray} 
respectively, where  $\lambda_{l_1}$ is the eigenvalue of
$V_{S,12}$ which 
can be readily calculated.
Using Eqs.(\ref{OPE}), (\ref{OPE2}), (\ref{calculus})
and the
summation formula
\begin{displaymath}
\sum_{k=0}^N
\frac{(-1)^kN!}{k!(N-k)!}M(k+1,1,-t)=\frac{e^{-t}t^N}{N!}
\end{displaymath}
we obtain the  expression for the 
eigenvalue of $V_{S,12}$ in the form of the integral
\begin{eqnarray}\label{lambdaL}
\lambda_{l_1}=
\int\limits_0^{\infty} r dr V(r)f_{l_1}(r) 
\end{eqnarray}
with $f$ the functions defined as
\begin{eqnarray}
f_{l_1}(r)=e^{-r^2/2}
\left[ \frac{r^{2l_1}}{2^{l_1}l_1!} - 1 + \frac{l_1}{4}\left(2 -
\frac{r^4}{4}\right) \right] .
\end{eqnarray}
The eigenvalue 
$(-1)^{l_1}$ of $\pi_{12}$
helps now to separate out the states with wrong symmetry:
the eigenvectors with $l_1$ odd are antisymmetric in $z_1,z_2$,
and the projector 
$P_S$
in (\ref{INERTIA}) 
eliminates their contributions.
All even values of $l_1$ ($\leq$$L$) 
can contribute to the bosonic sector, and the corresponding
$\lambda$'s must be checked.
Therefore, the set of inequalities
for the {\it control eigenvalues} 
\begin{equation}\label{criterion}
\lambda_{2n} \geq 0
\end{equation} 
for all values of  integer $n$ obeying $2$$n$$\leq$$L$
forms
the sufficient condition 
for (\ref{susy3}), i.e.,
for the 
``perturbation''
$V_S$ to be 
non-negative definite, criterion  {\bf (c)}
\footnote{Note that $f_{0}(r)\equiv 0$ and $f_{2}(r) \equiv 0$, 
therefore $\lambda_0=\lambda_2=0$.}
Under these conditions,  the triad 
$V_0$, $V_S\equiv\tilde{V}-V_0$ 
and $\{ {\cal P}_v|L\rangle\}$
obeys the 
``algebraic decomposition''
with properties {\bf (a)},  {\bf (b)} and  {\bf (c)},
with ${\cal P}_v|L\rangle$ being the 
ground state
in its sector $L,v$ with the energy ${\cal E}_{min}$
equal to eigenvalue of $V_0$.

The conditions (\ref{criterion}) can be used to describe the universality 
class of the interactions $V(r)$. They are analyzed explicitly
in the following sections.

\section{Generalized Yrast States and their Spectra}
\label{sec:level8}

With the criterion (\ref{criterion}) and 
Eqs.(\ref{multivortex2}) and (\ref{V0})
at hand, we can formulate very general result:
For any 
{\it bona fide} two-body
potential $V(r)$ which satisfies the integral condition
\begin{eqnarray}\label{RESULT1}
\int\limits_0^{\infty} V(\sqrt{2t}) e^{-t}
\left[ \frac{t^{2n}}{(2 n)!} - 1 + n\left(1 -
\frac{t^2}{2}\right) \right] \geq 0  
 \nonumber\\
\qquad for \quad 
any \quad n\leq L'/2,
\end{eqnarray}
the eigenstates of the Hamiltonian 
$\tilde{H}$
{\bf with minimal
energies at 
given pair of $v$ and $L(\leq\min\{L',N\})$}
have universal form, as follows from (\ref{multivortex2}),
\begin{eqnarray}\label{RESULT2}
|L, v ) = e^{-\frac{1}{2}\sum|z_i|^2} Z^v 
\left( \frac{\partial}{\partial Z}\right)^{N-L+v}
\prod\limits_{k=1}^{N} (z_k - Z), 
\qquad
 \nonumber\\
\qquad
Z\rightarrow\frac{1}{N}\sum\limits_{i=1}^N z_i .
\end{eqnarray}
Their energies are given by (\ref{V0}) and they can be expressed 
through the simple moments of 
$V(r)$,
\begin{eqnarray}\label{RESULT3}
{\cal E}_{min}(L,v)= L + N + \frac{N(N-1)}{2} s_0
+ \frac{ s_0 - s_1 }{2} N (v-L)
\\
s_0=\int_{0}^{\infty}dt e^{-t}V(\sqrt{2t}), \quad
s_1=\int_{0}^{\infty}dt e^{-t}
\left(\frac{1}{2}+\frac{t^2}{4}\right)V(\sqrt{2t})\nonumber
\end{eqnarray}
which
are equal to  
expectation values of interaction between
two bosons both in the ground 
$s_0=
\langle \psi_{0}^{\dagger} \psi_{0}^{\dagger} |
V |  \psi_{0} \psi_{0} \rangle$ 
and the first excited state
of oscillator 
$s_1=
\langle \psi_{1}^{\dagger} \psi_{1}^{\dagger} |
V |  \psi_{1} \psi_{1} \rangle$ , 
respectively.
Here, $\langle \psi^{\dagger} \psi^{\dagger} |
V |  \psi \psi \rangle$ denote  the usual two-body matrix
element and $\psi_{0}=  
|0\rangle   $  and  $ \psi_{1}= z  |0\rangle  $.
It is worth to note that this relation does not 
mean that the eigenvalues 
(\ref{RESULT3}) can be obtained as
an expectation value over a simple state, 
say, 
with the two lowest oscillator levels occupied; in fact,
the eigenstates (\ref{RESULT2}) 
are much more complicated 
when written
in the second quantization representation. 

At fixed $L$, we have exactly $L$
such 
equidistant {\it generalized yrast}
states, marked by $v=0,1,2,...,L$ ($v\neq L-1$),
Cf. 
Fig. 3., 
right hand part.
Each such state is the ``ground state'' in the sector $L,v$
(of course, there are other
states in each sector with higher energies).

Example of the spectrum of a real system 
for $N=L=6$
and the $\delta$-interaction
 is shown in  
Fig. 5.

The usual yrast states are those of (\ref{RESULT2}) that minimize
${\cal E}_{min}(L,v)$ 
with respect to $v$, Cf. Eq.(\ref{YRASTmin}).
As ${\cal E}_{min}(L,v)$ in
(\ref{RESULT3}) depend linearly on $v$, it is immediately seen 
that there is only two
options
\footnote{
The rare case of exact equality in (\ref{OPTIONS2}) corresponds to
complete degeneracy of the generalized yrast states 
with different values of $v$, 
having the same $N$ and $L$, see, e.g. Ref.\cite{HV3}). }  
:

\begin{eqnarray} \label{OPTIONS1}
{\bf (I)} \qquad v=0, \qquad  if  \qquad  D=s_0-s_1  > 0 ,
\end{eqnarray}
\begin{eqnarray} \label{OPTIONS2}
{\bf (II)} \qquad v=L \qquad  if   \qquad D=s_0-s_1  \leq 0 .
\end{eqnarray}
where the ``spectral discriminant'', $D$, is given by 
\begin{eqnarray} \label{discriminant}
D = 
\int_{0}^{\infty}dt e^{-t}
\left(\frac{1}{2} - \frac{t^2}{4}\right)V(\sqrt{2t}) .
\end{eqnarray}
For the {\it bona fide} potentials obeying (\ref{RESULT1}),
the first option is usually realized,
as is assumed in 
Figs. 3 and 4, see also  Figs. 5 and 7 for
particular interaction potentials. 
For example,
integrating by parts in (\ref{discriminant}), 
the spectral discriminant $D$
is seen 
positive
for any decreasing potential 
\begin{equation} \label{condYRAST}
\frac{ dV(r)}{  dr} < 0 
\end{equation}
which allows the representation (\ref{OPE}).
The condition (\ref{condYRAST}) can be replaced by 
\begin{equation} \label{condYRAST1}
\frac{ dV(r)}{  dr} \leq 0 ,
\end{equation}
if strict inequality in (\ref{condYRAST}) holds for at least 
single value of $r$. 

The inequality 
$({\cal V}_0-{\cal V}_1) (v-L) < 0$ corresponding to {\bf (I)}   
means that internal rotational excitations 
with higher $J=L-v$
are energetically 
favorable, once the interaction energy
between two bosons
in the state $z$$|$$0$$\rangle$ is smaller than that in the state 
$|$$0$$\rangle$.

Physically, the {\it yrast} wave functions (\ref{RESULT2}) with $v=0$
correspond to condensation to a vortex, rotating around the
``center-of-mass'', as discussed, for example, in \cite{Wilkin}.

In contrast, the maximum seniority states with $v=L$,
which correspond to purely collective rotation\cite{Mottelson}
with no internal excitations, were shown\cite{Wilkin} to
be energetically favorable 
in the case of  
attractive $\delta$-forces.
It is curios that there exists a broad class of 
predominantly attractive interactions, for which the yrast
states have the same form as for the attractive $\delta$-function
interaction \cite{praCOM}.

The results desribed in this section are very general \cite{HVprl}.
The condition (\ref{RESULT1}) defines the class of the potentials
for which the results (\ref{RESULT2},\ref{RESULT3}) are valid.
In the next section, we show that  (\ref{RESULT1}) holds
for many potentials of physical interest and consider few explicit
examples. Discussion of general properties of (\ref{RESULT1})
will be given in section 10.

\section{The results for particular interactions}
\label{sec:level9}

We consider now few applications of the results obtained in general 
form in the previous section to particular cases of various 
potentials of interaction between the bosonic atoms.

{\it Gaussian forces 
with variable range and $\delta$-interaction.}

We start with considering the case of   
repulsive Gaussian interaction, 
\begin{eqnarray}\label{GAUSS}
V(r)= \frac{U_0  }{\pi R^2}e^{-r^2/R^2}, \qquad U_0 \geq 0,
\end{eqnarray}
with $U_0$ a nonnegative strength and $R$ the radius 
which can be varied from zero to infinity.
We obtain from Eq.
(\ref{RESULT1}) 
the control eigenvalues 
\begin{eqnarray}\label{LAMBDA-GAUSS}
\lambda_{2n} = \frac{U_0}{\pi(2+R^2)}
\left[\left(\frac{R^2}{2+R^2}\right)^{2n}-1+
4n\frac{1+R^2}{(2+R^2)^2}\right] .
\end{eqnarray}
From this equation, it can be easily 
seen that the non-negative definiteness
condition (\ref{RESULT1}) $\lambda_{2n}\geq 0$ 
is fulfilled for any $n$ irrespectively to the value of $R$,
see 
Fig. 6.
This can be easily proved by induction in $n$.
Indeed, we have from (\ref{LAMBDA-GAUSS})
taht $\lambda_{0}=\lambda_{2}=0$ and  
\begin{displaymath}
\lambda_{2(n+1)} - \lambda_{2n} =
\frac{4U_0(1+R^2)}{\pi (2+R^2)} \left[1 - \left(
\frac{R^2}{2+R^2}\right)^{2n}\right]
 > 0,
\end{displaymath}
therefore $\lambda_{2n} > 0$ for any $n \geq 2$.
The results given by Eqs.(\ref{RESULT2},\ref{RESULT3}) 
are therefore valid
for any $L\leq N$, and the spectrum of the generalized yrast states
for the case of the Gaussian potential
is given by 
\begin{eqnarray}\label{E-GAUSS}
{\cal E}_{min}(L, v) =L+N+
 \nonumber\\
\frac{U_0}{\pi(2+R^2)}
\left[N(N-1)/2 -  \frac{(1+R^2)}{(2+R^2)^2}N(L-v)  \right]
\end{eqnarray}
One can see 
that the usual yrast states correspond to $v=0$.

Taking the zero range limit, $R \rightarrow 0$, in Eqs.
(\ref{GAUSS},\ref{LAMBDA-GAUSS},\ref{E-GAUSS}),
we pass to the case of the $\delta$-function repulsive interaction 
\begin{eqnarray}
V = U_0 \delta ( \vec{r} ) =  U_0 \frac{\delta ( r )}{2 \pi r} .
\end{eqnarray}  
In this case, we obtain instead of (\ref{LAMBDA-GAUSS})
\begin{eqnarray}
\lambda_{l_1} = \frac{U_0}{2\pi}
\left[\delta_{l_1, 0} + \frac{l_1}{2}-1\right].
\end{eqnarray}
It is seen that 
while 
$\lambda$$_1$$=$$-$$1$$/$$2$, for any $l$$_1$$=$$2$$n$ even 
the condition 
$\lambda_{2n}\geq0$ (\ref{RESULT1}) holds,
Cf.
Fig. 6. 
The energies of the generalized yrast states 
(\ref{RESULT3}) 
are now given by
\begin{eqnarray}\label{Edelta}
{\cal E}_{min}(L, v) =L+N+\frac{U_0}{8\pi}N(2N-L+v-2).
\end{eqnarray}
Yrast states have $v=0$,
and ${\cal E}_{min}(L,0)$
agrees
with that obtained numerically \cite{BP}, see 
also \cite{NOV1},\cite{NOV2},\cite{NOV3}.

It is expedient to look at the full spectrum of the system.
At low $N$ and $L$, the problem can be diagonalized analytically.
Below, we present the results for the total angular momentum $L=6$ 
in the system of six particles 
interacting via repulsive $\delta$-function
interaction. The total number of states ( dimensionality of the
partition basis ) is $p(6) = 11$. 
The interaction 
energies $E_k(L,v)$ of the states are given by 
\begin{eqnarray}
E_{0}(6,6) = 15 \frac{U_0}{2\pi}  ,\qquad
E_{0}(6,4) =  12 \frac{U_0}{2\pi} ,\qquad
E_{0}(6,3) =   \frac{21}{2} \frac{U_0}{2\pi} ,
\qquad\qquad
\nonumber\\
E_{0}(6,2) =   9 \frac{U_0}{2\pi} ,\qquad
E_{1}(6,2) =  \frac{21}{2} \frac{U_0}{2\pi} ,\qquad
 \nonumber\\
E_{0}(6,1) =   \frac{15}{2} \frac{U_0}{2\pi} ,\qquad
E_{1}(6,1) =   \frac{81}{8} \frac{U_0}{2\pi} ,\qquad
\nonumber\\
E_{0}(6,0) =   \frac{U_0}{2\pi} 6 ,
\qquad
E_1(6,0) =   \frac{U_0}{2\pi}
\left[ -\eta +\frac{69}{8} 
+ \zeta \right] ,
\qquad\qquad\qquad
\nonumber\\
E_2(6,0) =   \frac{U_0}{2\pi}
\left[ -\eta +\frac{69}{8} 
 -  \zeta \right] ,
\qquad
E_3(6,0) =   \frac{U_0}{2\pi} 
\left( 2\eta +
\frac{69}{8} \right)  \qquad\qquad
\end{eqnarray}
with
\begin{displaymath}
\eta = \frac{\xi}{32} +\frac{21}{4\xi}, \quad
\zeta = i\frac{3\sqrt{3}}{2}\left( \frac{\xi}{48} -\frac{7}{2\xi}
\right),
\quad
\end{displaymath}
\begin{displaymath}
\qquad\xi =(324+12 i \sqrt {32199})^{1/3}
\end{displaymath}
Here, the index $k$ marks degree of excitation of the state
within the same $L,v$-sector: so $E_{0}(L,v)= E_{min}(L,v)$
correspond to the generalized yrast states
(\ref{RESULT3},\ref{Edelta}).
There are no excited states in the sectors $v=6,4$ and $3$,
the sectors $v=4$ and $5$ have one excited state each.
The sector $v=0$ has three excited states,
their energies are found by solving a cubic equation.
The spectrum is shown in 
Fig. 5.

{\it  Two- and three dimensional Coulomb forces}
  
Of special physical interest are the long-range forces of
Coulomb type. 
For the conventional Coulomb interaction 
\begin{eqnarray}
V=\frac{U_0}{r} , \qquad  U_0\geq 0
\end{eqnarray} 
we obtain in (\ref{RESULT1})
\begin{eqnarray}\label{lambdaC}
\lambda_{2n}=
U_0\sqrt{2\pi}\left(  \frac{5 n}{16}   -\frac{1}{2} +
\frac{\Gamma(2n+1/2)}{2\sqrt{\pi} (2n)!} \right) .
\end{eqnarray}
It is easily seen, using induction in $n$,  that the
inequalities $\lambda_{2n}\geq 0$ (\ref{RESULT1}) hold 
(Cf.
Fig. 6), 
and the wave functions 
of the generalized yrast states are therefore given by (\ref{RESULT2}).
Their energies are given by 
\begin{eqnarray}
{\cal E}_{min}(L, v) =
L + N + \frac{U_0\sqrt{2\pi}}{64} N [16(N-1) - 5(L-v)] .
\end{eqnarray}
The yrast states correspond to $v=0$.

Similar formulas can be obtained for 
the two-dimensional Coulomb log-interaction 
\begin{eqnarray}
V=U_0 log\left(\frac{1}{r}\right) , \qquad  U_0\geq 0 ,
\end{eqnarray}
which corresponds to repulsion at small distances.
We have
\begin{eqnarray}\label{lambdaLOG}
\lambda_{2n}=\frac{U_0}{4}
\left[3n-\frac{1}{n}+2\psi(1)-2\psi(2n)\right] \geq 0 ,
\end{eqnarray}
where $\psi$ is digamma function \cite{ABR},
the inequality in  $\lambda_{2n}\geq 0$ for (\ref{lambdaLOG}) 
can again be easily proven by induction,
Cf.
Fig. 6, 
so condition
(\ref{RESULT1}) holds.
The energies of the generalized yrast states 
are given by 
\begin{eqnarray}
{\cal E}_{min}(L, v) =
\nonumber\\
L+N-\frac{U_0 N}{4}\left[ (log(2)-\gamma)(N-1) 
-\frac{3}{4}(v - L) \right]
\end{eqnarray} 
with $\gamma=-\psi(1)=0.57721...$ the Euler constant.
The yrast states correspond to $v=0$.  

The results very similar to the above 
can be obtained for the screened Coulomb 
(Yukawa) forces $V \propto exp(r/r_0)/r$.

The results for the generalized yrast states in the cases 
considered in this section are illustrated in 
Fig. 7 
for $N=L=6$.

\section{The  sufficiency condition and the universality class
of bona fide repulsive potentials}
\label{sec:level10}

The simple sufficiency condition  (\ref{RESULT1})
can be checked straightforwardly for any interaction 
potential $V(r)$ of interest, as was done in the previous section.
It is interesting to understand why many potentials do meet
this condition.
In this section, we show 
that the applicability condition (\ref{RESULT1}) 
imposes in fact {\bf only weak}
restrictions
on the 
class of 
forces $V$$($$r$$)$
which can be regarded as predominantly repulsive.
We will discuss the condition (\ref{RESULT1}) in more details
and define some subclasses of the bona fide potentials of 
interest.

{\it Short-range interactions}

The factor functions $f_{2n}(r)$ in expression
$\int_0^{\infty} r d r f_{2n} (r)U(r)$
of Eq. (\ref{RESULT1})
are plotted in 
Fig. 8. 
At small $r\simeq 0$, 
the factor-functions 
approach positive constant values with zero derivatives,
\begin{equation}\label{limit}
f_n\Big{|}_{r=0}=n-1 , \qquad  \frac{df_n}{dr}  \Big{|}_{r=0}=0, 
\end{equation}
while the first node of the functions $f_{2n}(r)$ 
occurs at 
\begin{eqnarray}\label{r0}
r_0 = \sqrt{2\sqrt{12-2\sqrt{30} } } \simeq1.43
\end{eqnarray}
which is of order of the oscillator length,
$\frac{\hbar}{m\omega}$,
in our units ($\hbar=m=\omega=1$).
From Eqs.(\ref{limit}) and (\ref{r0}), it follows therefore that
for any short range
(as compared to the characteristic length of the trap)
interaction, the condition (\ref{RESULT1}) 
reduces to the single inequality
\begin{eqnarray} \label{INTEGRAL}
\int d^2 \vec{r} V(r) \geq 0,  
\end{eqnarray}
Thus, the results (\ref{RESULT2},\ref{RESULT3})
hold for
short-range interactions, which 
are repulsive on average. 
I is seen also that inequality (\ref{INTEGRAL}) gives $v$$=$$0$ for
the usual yrast states, according to (\ref{OPTIONS1}).

It is interesting that the sort-range potentials do not have
to be purely repulsive to match the condition (\ref{RESULT1}),
(\ref{INTEGRAL}). Instead, the  condition (\ref{RESULT1}) implies 
that $V(r)$ has sufficiently strong repulsive
component.
For example, consider the two-parameter family of the potentials 
in the form 
\begin{eqnarray}\label{Ubound}
V ( r ) = - \frac{| V _0 |}{R^2} 
( 1 - ( r / R ) ^2 ) e ^{-\frac{r^2}{2R^2} }
\end{eqnarray}
These potentials can be regarded as attractive in the usual sense, 
having the potential well at short distance.
Examples are shown on 
Fig. 9. 
In fact, if the 
well is deep enough, it can even support
bound state.  Indeed, consider the Schroedinger equation
of relative motion ($s$-wave) of the pair of particle with unit mass
each, interaction via (\ref{Ubound}),
\begin{displaymath}
- \frac{1}{r} \frac{d}{dr}\left( r \frac{d}{dr} \right) \phi_0 +
V(r)\phi_0
 = \varepsilon_0 \phi_0,
\end{displaymath}
for the wave function of the ground state, $\phi_0$.
We obtain rigorous variational upper bound on the ground state
energy, $\varepsilon_0$,
using the trial wave function $\phi_0 = exp(-\sqrt{3}r^2/R^2)$:
\begin{displaymath}
\varepsilon_0 < \frac{ \sqrt{6V_0}\left[    
720 V_0^2+1
     -8 (6V_0)^{5/2}/3-46V_0-2(6V_0)^{3/2}
     \right]}{R^2(24V_0-1)^2}
\end{displaymath}
It is seen that we have negative energy of the ground state, 
$\varepsilon_0 < 0$,  if 
\begin{eqnarray}\label{Ubound1}
\qquad V_0 >
  \nonumber\\
\frac{1}{216}\left[ (3763+18\sqrt{502})^{1/3}
+ \frac{241}{ (3763+18\sqrt{502})^{1/3} } + 13
     \right]^2
  \nonumber\\
 \simeq 8.99
\end{eqnarray}
in such cases the potential well indeed supports the bound state(s).
On the other hand, the potential  (\ref{Ubound})
satisfies the basic condition (\ref{RESULT1}) if 
\begin{eqnarray}\label{Ubound2}
R \leq \frac{1}{10}\sqrt{5+5\sqrt{41}}  \simeq  0.608, 
\end{eqnarray}
with all its control $\lambda_{2n}$ nonnegative,
irrespectively of the magnitude of $V_0$,
Cf. 
Fig. 9, lower panel.
Therefore, the family of the potentials (\ref{Ubound})
satisfying  (\ref{Ubound1}) and  (\ref{Ubound2})
will be ``predominantly repulsive'' in the sense of the
condition (\ref{RESULT1}), having the solutions
for the generalized yrast states in the form (\ref{RESULT2}) and
 (\ref{RESULT3}), while having strong attractive 
 component sufficient for the bound state.
In the sense of three-dimensional scattering theory,
the scattering length can  be negative.

{\it Long-range forces}

The condition (\ref{RESULT1}) holds even 
for many long-range interactions.
This is seen from behavior of function $f_4(r)$
($n=0$ and $n=1$ give $f\equiv0$)
which is positive at $r<r_0$ and 
\begin{eqnarray}
r > r_1 = \sqrt{2\sqrt{12+2\sqrt{30} } } \simeq 3.10 ,
\end{eqnarray}
and $f_4$ is negative only in the interval $r_0 \leq r \leq r_1$
($f_n(r)$ for higher $n$ behave similarly).
By direct calculation, 
it is easy to see that the integrals of the functions $r f_{2n}$
[Cf. Fig. 8, lower panel]  
over $r$ vanish, 
\begin{eqnarray} \label{AREAS}
\int\limits_0^{\infty}rdr f_{2n}(r) =
S_1  + S_2 + S_3 = 0 , 
\qquad \qquad \qquad \qquad 
\nonumber\\
S_1 = \int\limits_0^{r_0}rdr f_{2n}(r) ,  
\quad
S_2 = \int\limits_{r_0}^{r_1}rdr
f_{2n}(r) ,
\quad
S_3  =   \int\limits_{r_1}^{\infty}rdr
f_{2n}(r),
\end{eqnarray}
so the net areas coming form the 
regions where $f>0$ and $f<0$ coincide,
\begin{displaymath}
S_1 + S_3 = |S_2|,
\end{displaymath}
as is shown in 
Fig. 10,
upper panel, for the case $f_{4}$.
From this geometry, which is also illustrated in 
Fig. 10,
lower panel,
it is clear that (\ref{RESULT1}) holds, if $V(r)$ 
decreases monotonically and fast enough,
as is the case for 
the long range Gaussian
and Coulomb forces.
Example of the potentials of such type are shown in 
Fig. 10,
lower panel.

Integrating by parts, one sees that (\ref{RESULT1}) 
holds for any interaction described by monotonically decreasing 
and concave function $V(r)$
\footnote{It is assumed that the potential $V$ can be represented
in the form of operator expansion (\ref{OPE}).},
\begin{eqnarray}  \label{MONOTONIC}
\frac{dV(r)}{  dr} \leq 0 , \qquad 
\frac{d^2 V(r) }{  dr^2} \geq 0 , 
\end{eqnarray}
not necessarily repulsive everywhere.
This means, in particular, that the solutions (\ref{RESULT1}) 
apply to the exponential and screened Coulomb potentials,
\begin{eqnarray}
V_{e}(r) = |V_0| exp(-r/R), \qquad 
V_{sc}(r) = |V_0| \frac{exp(-r/R)}{2\pi r R} .
\end{eqnarray}

One can therefore summarize that
the condition holds for any physically meaningful repulsive 
interaction.

\section{Conclusion}
\label{sec:level11}

To conclude, we considered the problem of weakly interacting
Bose atoms in symmetric harmonic trap.
Our main purpose was  to study the yrast states of the
system.
In order to treat the problem of the ground state 
of the system at a given angular momentum, 
we developed an expansion of the 
interaction in powers of ladder operators.
This universal operator expansion is exact and it provides a very 
convenient way to study the eigenvalue problem in 
the coordinate representation. 

Taking into account the additional conserved quantity,
associated with the collective contribution to the
total angular momentum,
we considered a more general problem, namely, the
ground state of the system as a function of two
conserved quantum numbers: the total angular momentum and
the angular momentum of internal excitations.
We called this series of states ``generalized yrast states''.

A method of ``algebraic decomposition''
for the interaction 
has been developed to derive analytically
the states with minimum energy at a given angular momentum
and seniority (generalized yrast
states). The results apply, in particular, to the usual yrast states. 
The wave functions of the generalized yrast state have the
form of
``condensed vortex states''. Their energies are expressed
through their quantum numbers and 
simple integrals 
of the 
interaction potential. 

We studied the condition on the interaction potential
which allows the use of our solution.
Analysis shows that there exists 
a broad {\it universality} class of the repulsive 
interactions for which these results are valid. 
We described this universality class by simple integral 
condition on the interaction potential.

The results of the work allow further generalizations and 
developments.
The three-dimensional case
can be done using the same method.
It is also interesting to study region of higher angular momenta
$L>N$, where the structure of the basis polynomials will
be changed \cite{symmetricFUNCTIONS}, while the numerical studies
indicate signs of phase transition\cite{BP}.
The method of 
``algebraic decomposition''
developed here is not
restricted to this particular problem and can be applied to 
fermions and even to the particles with parastatistics.

The work was supported by FAPESP.

\newpage
{\large Figure Captions}

\vspace{5mm}
Fig. 1.  
Definition of the yrast states.
The spectrum of a system (circles) is shown versus
the angular momentum, $L$. The lowest energy states
at each value of $L$, connected by the curve,
compose the ``yrast line'' of the system.

\vspace{5mm}
Fig. 2.  
Fragment of the spectrum of the system (schematic plot)
within approximation of the weak coupling limit.
The left hand side is the spectrum of the system
without interactions.
The levels are equidistant with the spacing $\hbar \omega$, 
each level is $p$-fold degenerate.
Central part is the spectrum of interacting system.
Each $L$-level splits onto $p$ levels. The sequence of
the states with minimum energy at given $L$ is the
yrast-line (right hand part).

\vspace{5mm}
Fig. 3.  
Illustration to the definition of the generalized yrast
states. The left-hand side is the part of the 
spectrum of the interacting system with 
definite angular momentum $L$ (corresponding to a bunch of
levels in the central part of Fig. 1.). 
These states can be grouped
onto ``bands'' with definite values of the seniority, $v$,
as is shown on the right hand side.
The sequence of lowest energy states in their $v$-sector
composes the generalized yrast states (they are marked with
asterisks). One of those, with absolute minimum of energy,
is the usual ground state.

\vspace{5mm}
Fig. 4.  
Illustration of 
``algebraic  decomposition''.
The spectrum of $V_0$ (left) is sequence of 
degenerate levels, labeled by the conserved
quantum number $v$. 
The non-negative definite
perturbation $V_S$$\geq$$0$ splits each level,
pushing the states up and leaving the lowest energy
in each 
$v$-sector  intact. 
The resulting spectrum of $V_0+V_S$
is shown on the right.

\vspace{5mm}
Fig. 5.  
The spectrum of the system with repulsive $\delta$-interaction
calculated analytically for the case of six particles in
the sector with total angular momentum $L=6$. 
The total number of states is $p(6)=11$.
The energy of the levels
$e(L,v)=\frac{2\pi}{U_0}[{\cal E}(L,v)-L-N]$, Eq.(\ref{Edelta}), 
are plotted against the internal angular momentum, $J=L-v$.
The generalized yrast states drop on the straight line
described by Eq.(\ref{Edelta}), they are connected by dashed
line.

\vspace{5mm}
Fig. 6.  
``Redge trajectories'' for $\lambda_{l_1}$ (\ref{lambdaL}).
Upper panel: the values of $a=\frac{2\pi}{U_0}\lambda_{l}$ are plotted
versus $l$ (circles connected by curves) 
for two different interaction potentials.
The curves  
correspond to $l$ considered as continuous variable.
Solid curve - repulsive Gaussian potential with $R=1$ and dashed curve -
repulsive $\delta$-function interaction.
The double circles denote the control eigenvalues with even
$l=2n$, Eqs.(\ref{RESULT1},\ref{LAMBDA-GAUSS}).

Lower panel: the same but for the Coulomb interaction (solid curve)
and $log$-Coulomb interaction (dashed curve), 
Cf. Eqs.(\ref{lambdaC},\ref{lambdaLOG}).

\vspace{5mm}
Fig. 7.  
The specta of the generalized yrast states
$e = \frac{2\pi}{U_0}{\cal E}_{min}(L,v)$
are plotted against $J=L-v$ ($N=L=6$) for the
Gaussian interaction with $R=1$ (small circles connected by solid line),
$\delta$-interaction (small circles, dashed line),
Coulomb interaction (big circles, solid line) and the $log$-Coulomb
interaction (big circles, dashed line).

\vspace{5mm}
Fig. 8.  
Upper panel:
Factor-functions $f_{2n}(r)$ for
$n=2, 3, 4, 5$ are plotted against $r$ :  $f_4$ (curve $1$), 
 $f_6$ (curve $2$),  $f_8$ (curve $3$) and  $f_{10}$  (curve $4$).

Lower panel: the same as above but for the $r f_{2n}(r)$.

\vspace{5mm}
Fig. 9.  
Upper panel: the potentials [Eq.(\ref{Ubound})] supporting the bound state 
with $V_0=10$ obeing Eq.(\ref{Ubound})
are plotted against $r$ for three values of $R$:
$R=0.3$ - solid curve,   $R=0.6$ - dashed curve, and   $R=1.0$ -
dashed-dotted curve.    

Lower panel: the ``Redge trajectories'' for $\lambda_{l_1}$
(\ref{lambdaL}). The eigenvalues $\lambda_l$ in combinations
$a=\frac{1}{U_0}\lambda_{l}$ 
are plotted as function of $l$
(circles) for the potentials shown on the upper panel. 
The curves connecting the symbols 
correspond to eigenvalues as continuous function
of $l$: $R=0.3$ - solid curve,   $R=0.6$ - dashed curve, and   $R=1.0$ -
dashed-dotted curve.   
The double circles denote the control eigenvalues with even
$l=2n$.

\vspace{5mm}
Fig. 10.  
Upper panel: areas under curve $r f_{4}(r)$ (illustration to 
Eq. (\ref{AREAS}).

Lower panel: the function  $r f_{4}(r)$ (thin curve) is plotted against $r$
together with potentials matching  condition (\ref{RESULT1}).
The latter are shown by
thick curves:
$1$ - Gaussian potential $V(r)= \frac{U_0}{\pi R^2} exp(-r^2/R^2)$
with $R=U_0=1.5$, $2$ - Coulomb potential  $V(r)= \frac{U_0}{r}$ with
$U_0=1/5$ and  $3$ - log-Coulomb potential  $V(r)= U_0 log(1/r)$ with 
$U_0=1/3$.


\begin{thebibliography}{300}  


\bibitem{ANDERSON} M.H. Anderson, J.N. Ensher, M.R. Matthews,
C.E. Wieman,
and E.A. Cornell, {\it Science}, {\bf 269} (1997), 198.

\bibitem{nature} D.A.Butts and D.S. Rokhsar, 
{\it Nature} {\bf 397} (1999), 327.

\bibitem{Wilkin}
N. K. Wilkin, J. M. Gunn, and R. A. Smith,
{\it Phys. Rev. Lett.} {\bf 80} (1998), 2265.

\bibitem{Mottelson}
B. Mottelson, {\it Phys. Rev. Lett.} {\bf 83} (1999), 2695.


\bibitem{BP}
G. F. Bertsch and  T. Papenbrock, {\it Phys. Rev. Lett.}  
{\bf 83} (1999), 5412.

\bibitem{REVIEW}
F. Dalfovo, S. Giorgini, L. P. Pitaevskii, and S. Stringari,
{\it Rev. Mod. Phys.} {\bf 71} (1999), 463;
P. Coullet and N. Vandenberghe, {\it Phys. Rev. E}, {\bf 64} (2001),
025202.

\bibitem{Matthews}
M.R. Matthews, B.P. Anderson, P.C. Haljan, D.S. Hall, C.E. Wieman, 
and E.A. Cornell, 
{\it Phys. Rev. Lett.} {\bf 83} (1999), 2501.

\bibitem{MarzlinZhangWright}
K.-P. Marzlin  W. Zhang and E. M. Wright, 
{\it Phys. Rev. Lett.} {\bf 79} (1997), 4728;
K.-P. Marzlin and  W. Zhang
{\it Phys. Rev. A} {\bf 57} (1998), 4761.

\bibitem{A} 
M. Linn, M. Niemeyer, and A. Fetter, 
{\it Phys. Rev. A} {\bf 64}  (2001),  023602;
J. R. Abo-Shaeer, C. Raman, and W. Ketterle,
{\it Phys. Rev. Lett.} {\bf 88} (2002), 070409.

\bibitem{BM} A. Bohr and B. Mottelson, ``Nuclear Structure'', 
Benjamin, New York, 1969.

\bibitem{FESHBACH} 
E. Timmermans, P. Tommasini, R. Cote, M. Hussein, A.K. Kerman,
{\it Phys. Rev. Lett.}  
{\bf 83} (1999), 2691;
E. Timmermans, P. Tommasini, M. Hussein, A.K. Kerman, 
{\it Phys. Rep.} 
{\bf 315} (1999), 199.

\bibitem{STRENTHmanipulating}
J. L. Roberts, N. R. Claussen, S. L. Cornish, E. A. Donley, 
E. A. Cornell, and C. E. Wieman, Phys. Rev. Lett. {\bf 86}, 4211 (2001). 

\bibitem{Laughlin} 
R.B. Laughlin, {\it Phys. Rev. Lett.} {\bf 50} (1983), 1395.

\bibitem{Trugman-Kivelson}
S.A. Trugman and S. Kivelson, 
{\it Phys. Rev. B} {\bf 31}  (1985),  5280.


\bibitem{HV1HV2}
M.S. Hussein and O.K. Vorov,
{\it New results for the yrast spectra of 
weakly interacting identical bosons},  
June 2000, Abstr. subm. to the
Int. Symposium {Nuclei and Nucleons 2000}, Darmstadt;
{\it Analytical results for the yrast spectra of 
interacting bosons}, 
July 2000, Abstr. submitted to the
Annual Meet. of the Brazilean Phys. Society.



\bibitem{NOV1}
R.A. Smith and N.K. Wilkin,
{\it Phys. Rev. A} {\bf 62} (2000), 061602.

\bibitem{NOV2}
A.D. Jackson and G.M. Kavoulakis,
{\it Phys. Rev. Lett.} {\bf 85} (2000), 2854.

\bibitem{NOV3}
T. Papenbrock and G.F. Bertsch, 
{\it J. Phys. A} {\bf 34} (2001), 603. 

\bibitem{dob1} W.J. Huang,   {\it Phys. Rev. A}  {\bf 63} (2001), 015602.


\bibitem{dob2} T. Nakajima and M. Ueda,
{\it Phys. Rev. A},
{\bf 63} (2001), 043610.    

\bibitem{dob3}
T. Papenbrock and  G. F. Bertsch,  {\it Phys. Rev. A}   
{\bf 63} (2001), 023616.

\bibitem{HVprl} M.S. Hussein and
O.K. Vorov,
{\it Phys. Rev. A} {\bf 65} (2002), 035603. 

\bibitem{physicaB} M.S. Hussein and O.K. Vorov,
{\it Physica B} {\bf 312-313} (2002), 550; 
reported at the SCES-2001 Conference
at Ann Arbor, 2001.

\bibitem{HV3} M.S. Hussein and
O.K. Vorov,
subm. to {\it Phys. Lett. A.}

\bibitem{praCOM} M.S. Hussein and O.K. Vorov,
{\it Phys. Rev. A.} {\bf 65} (2002), 000, in press.

\bibitem{symmetricFUNCTIONS}
B.  Higman,  ``Applied group-theoretic and matrix methods'', 
New York, Dover, 1964. 

\bibitem{ABR} M. Abramowitz and I.A. Stegun, 
``Handbook of Mathematical 
Functions'', National Bureau of Standards, 1964.

\bibitem{SUSYprimer} There is a tiny formal point in common with 
``factorization method'', E.Schr{\"o}dinger, 
{\it Proc.R.Irish.Acad. A} {\bf 46} (1940), 9, and 
some ideas used in  supersymmetry,
E.Witten, {\it Nucl. Phys. B} {\bf 188} (1981), 513;
F.Cooper, J.N.Ginocchio, and A.Khare, 
{\it Phys. Rev. D} {\bf 36} (1986), 2458
(we note that the non-negative perturbation
has multiply degenerate ground state). 

\bibitem{BOOK-MATRICES} G.W.Stewart and J.-G. Sun, 
``Matrix Perturbation Theory'', Academic Press, New York, 1990.

 
\bibitem{COMMENTdegeneracy} In principle, this does not exclude 
possibility that
the ground state can still be degenerate, 
if it was such
for $V_0$. 
It is possible to show\cite{HV3} that
in our case 
the ground states of $\tilde{V}$ are not degenerate.



\end{thebibliography}
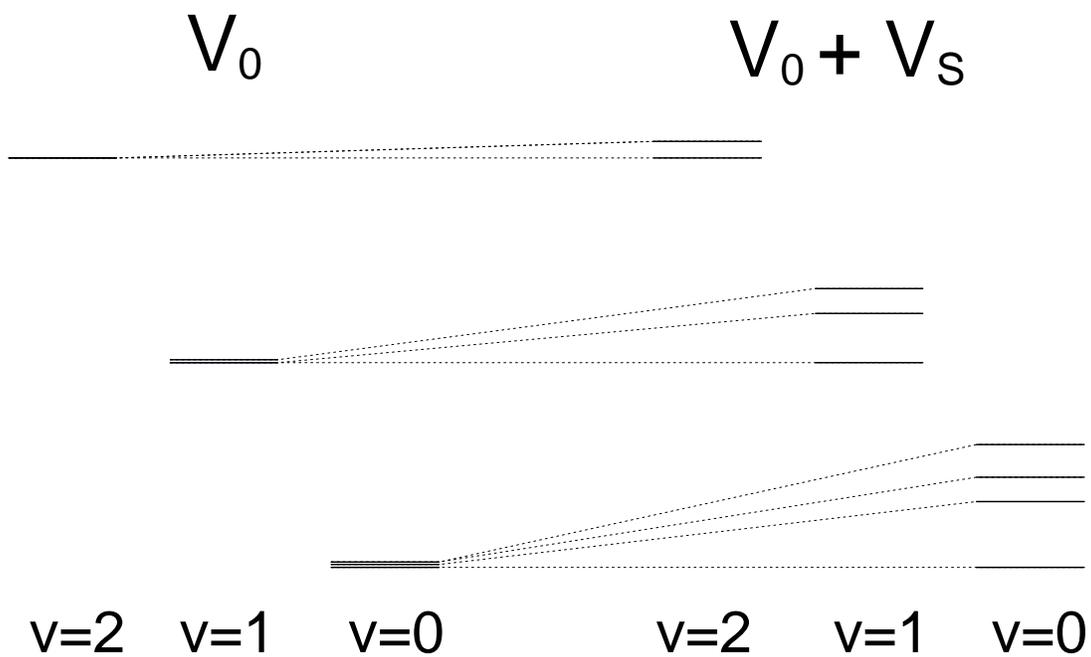
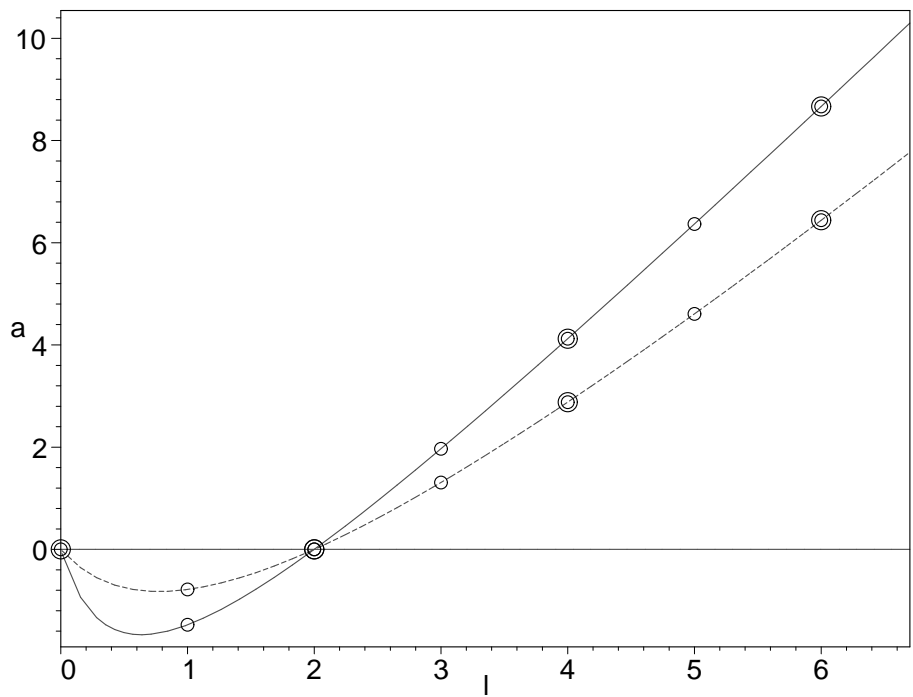
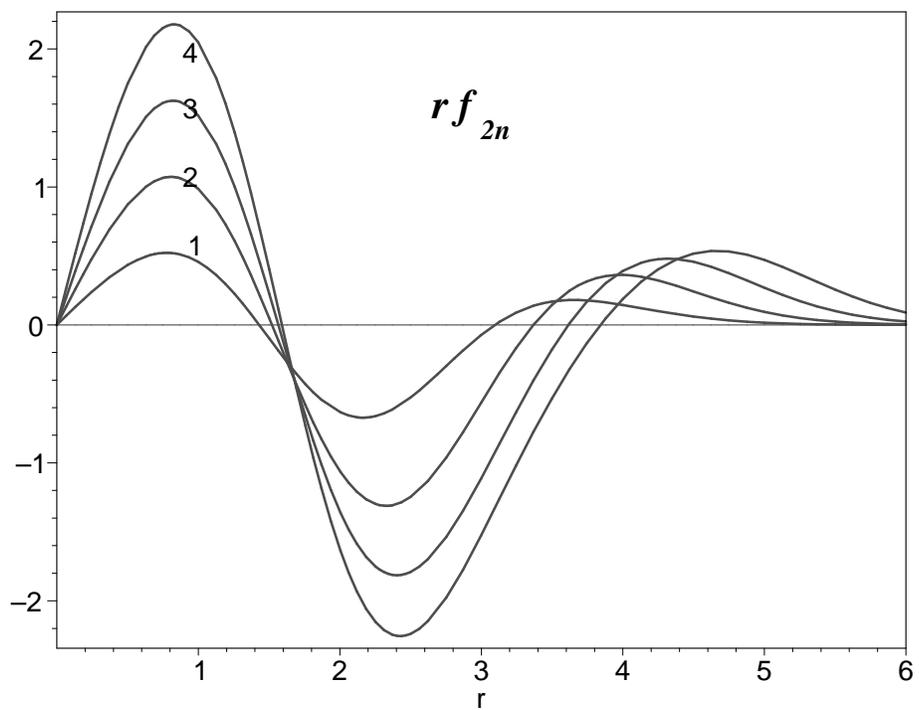
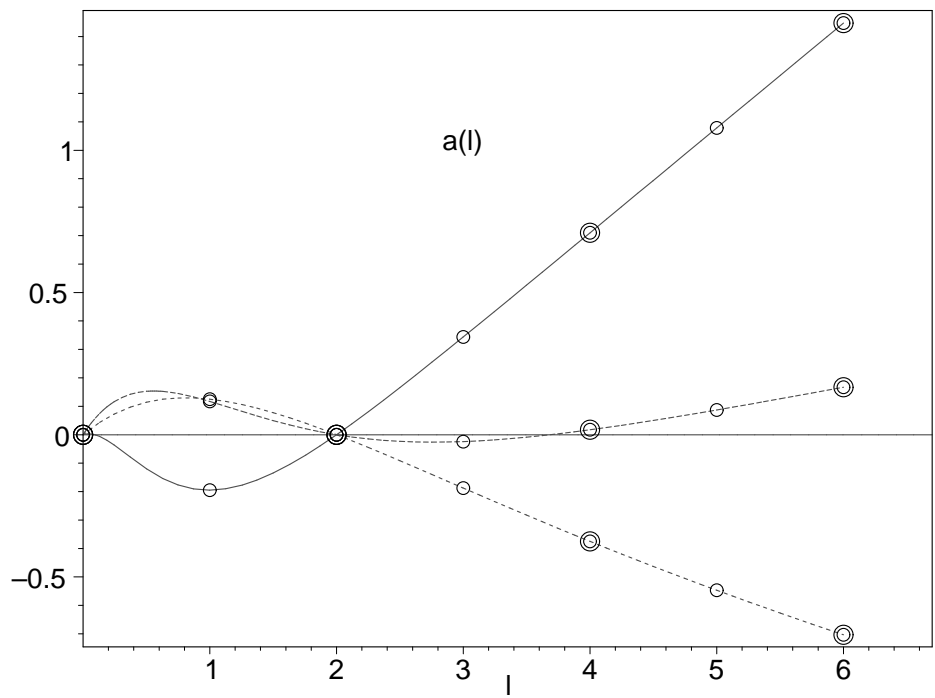
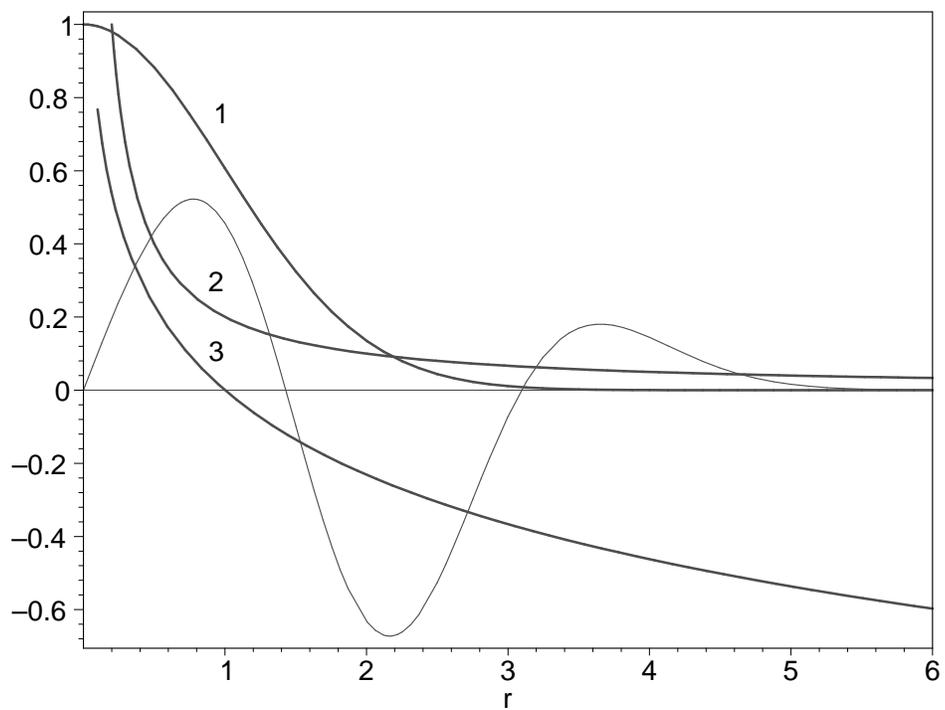
\end{document}